# Scientific Paper Classification Based on Graph Neural Network with Hypergraph Self-attention Mechanism

Jiashun Liu, Zhe Xue*, Ang Li

[1]School of Computer Science, Beijing Key Laboratory of Intelligent Telecommunication Software and Multimedia, Beijing University of Posts and Telecommunications, Beijing 100876, China

**Abstract**：The number of scientific papers has increased rapidly in recent years. How to make good use of scientific papers for research is very important. Through the high-quality classification of scientific papers, researchers can quickly find the resource content they need from the massive scientific resources. The classification of scientific papers will effectively help researchers filter redundant information, obtain search results quickly and accurately, and improve the search quality, which is necessary for scientific resource management. This paper proposed a science-technique paper classification method based on hypergraph neural network(SPHNN). In the heterogeneous information network of scientific papers, the repeated high-order subgraphs are modeled as hyperedges composed of multiple related nodes. Then the whole heterogeneous information network is transformed into a hypergraph composed of different hyperedges. The graph convolution operation is carried out on the hypergraph structure, and the hyperedges self-attention mechanism is introduced to aggregate different types of nodes in the hypergraph, so that the final node representation can effectively maintain high-order nearest neighbor relationships and complex semantic information. Finally, by comparing with other methods, we proved that the model proposed in this paper has improved its performance.

**Keywords:** Hypergraph; Graph neural network; Graph attention;

## 1. Introduction

Graph neural network [1] has been well studied by researchers. The main methods only focus on homogeneous networks with only one kind of nodes and edges. However, many networks in the real world contain different types of nodes and edges, which constitutes heterogeneous information networks. Heterogeneous information network representation learning is usually based on meta paths [2]. For the task of classifying nodes of scientific papers, the meta path can be constructed as "author-paper-author", which represents the relationship between two authors publishing the same paper. We propose to build hyperedge relationships based on high-order subgraphs in the network. Compared with the original path, high-order subgraphs can better maintain the structural characteristics and semantic information of the network [3-6]. Therefore, hyperedges established by higher-order subgraphs maintain the heterogeneity of the network and distinguish the relationships between different types of nodes, which are crucial in the process of learning the representation of heterogeneous information networks. In addition, hyperedges can effectively explore the high-order adjacency relationship in the network, and distinguish different semantic roles in the subgraph, so as to obtain a more complete semantic representation of the nodes in the subgraph.

On the other hand, the relationship between nodes in heterogeneous information networks is complex, such as affiliation and high-order interaction, which are far more than the point-to-point relationship between nodes in homogeneous networks [7]. In the network of scientific papers, authors can publish different papers at different meetings. If the relationship between papers is regarded as an edge in the graph, then the relationship of all papers of the same author can be considered as an edge, and this edge is not only connected to two nodes. The advantage of the heterogeneous hypergraph network [8] is that the heterogeneous attributes in the network can be expressed by introducing high-order structure information and fusing hyperedge forms formed by different types of high-order subgraphs. Therefore, introducing high-order relational structure information into the network is helpful to study heterogeneous information networks [9] from the perspective of hypergraph theory. After modeling heterogeneous information networks into hypergraph networks, graph neural networks are extended to heterogeneous hypergraph networks by using the basic properties of hypergraphs, so as to realize the node classification [10] of heterogeneous information networks. The main contributions of this paper are as follows:

With the help of high-order subgraphs, the heterogeneous information network is modeled as a heterogeneous hypergraph network, which retains the high-order nearest neighbor network information and complex semantic information in the heterogeneous information network, and propagates the node characteristics through the hypergraph convolution network to learn the network node representation.

A new attention mechanism for heterogeneous hypergraph networks of scientific papers is proposed, which can dynamically aggregate nodes with different semantic roles according to different importance, so that the model can obtain better node representation even in sparse large-scale heterogeneous information networks. At the same time, attention mechanism is introduced to dynamically allocate weights to nodes with different semantic roles, which also provides a meaningful explanation for the information aggregation process of nodes, and helps to make the decision-making process of heterogeneous information network analysis more transparent.

*Corresponding author: Zhe Xue (xuezhe@bupt.edu.cn).

The experimental results show that this method has achieved good performance in science-technique paper classification task, and greatly improved the training speed.

## 2. Related Work

Scientific paper classification belongs to the field of text classification in natural language processing. The commonly used text feature selection methods include singular value decomposition, word frequency reverse file frequency, document topic model and so on. For example, Giannis [11] and Wei et al. [12] proposed a paper classification method based on SVD, which uses SVD to extract the features that can best represent the paper category; Jacob et al. [13] and Li et al. [14] mine the topic probability distribution of patent text based on LDA, and use the topic similarity to classify the text; In the machine learning method based on Feature Engineering [15][19] the features obtained by SVD, TF-IDF and LDA are difficult to express the deep semantic features of this paper. With the development of deep learning technology, researchers gradually adopt the deep learning model based on word vector to classify this paper. Compared with the traditional word2vec [20][21], these pre- training models consider the context and solve the problem of polysemy; On the other hand, different levels of semantic features are obtained through hierarchical learning, which provides rich feature selection for downstream tasks.

On homogeneous networks, Tapas et al. [22] and Meng et al. [23] defined graph convolution based on spectral graph theory by introducing the Fourier transform of graphs. On this basis, Sui et al. [24] and Li et al. [25] used the Chebyshev polynomial of the eigenvalue diagonal matrix to approximate smooth the convolution kernel in the spectral domain. Hamilton used neural network to summarize the characteristic information of neighbors, and made the calculation not need to be completed on the whole graph through sampling strategy. Sun et al. [26] and Li et al. [27] used the node self-attention mechanism to measure the influence of different neighbors, and combined their influence to obtain the node representation. At the same time, in order to integrate high-order structural information when learning node representation, hypergraph representation learning has attracted attention in recent years. Wang et al. [28] and Kou et al. [29] proposed a hypergraph representation learning framework to capture the correlation between higher-order data by designing hyperedge convolution operations. Yadati et al. [30] and Li et al. [31] proposed a method of training GCN on hypergraph based on the spectral theory of hypergraph, so that nodes on the same edge have similar representation, which is used for semi supervised learning and exploring combinatorial optimization problems on hypergraph. While defining hypergraph convolution, Yang et al. [32] and Liang et al. [33] used the attention mechanism to measure the interaction between nodes in the hyperedge to obtain a better node representation. The above work only focuses on homogeneous networks. The way to construct hypergraphs is to model the first-order nearest neighbors of nodes or K nearest neighbors based on KNN algorithm as hyperedges. For heterogeneous networks, how to effectively establish hyperedges is still a problem worth studying.

## 3. The Proposed Method

The proposed scientific paper classification method SPHNN includes three main modules. First, hypergraph construction is based on high-order subgraphs, and heterogeneous information networks are constructed as multiple hypergraphs. Second, the hyperedge attention mechanism is introduced into the hypergraph convolution process to aggregate nodes with different semantic roles. Finally, through the attention mechanism, the node representations with different semantics based on multiple hypergraphs are fused to obtain the final comprehensive semantic representation. The whole structure of the proposed method is shown in Figure 1.

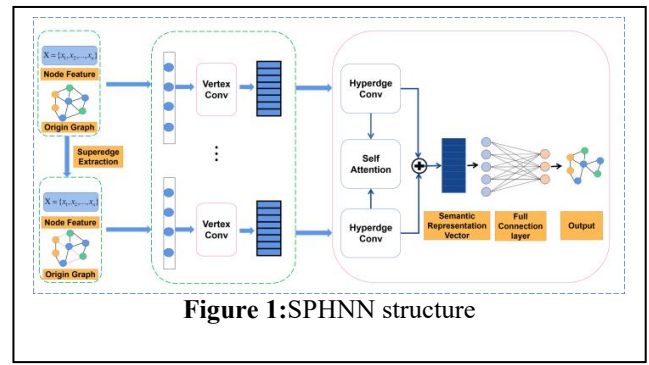

**Figure 1:** SPHNN structure

### 3.1 Hypergraph construction based on higher-order subgraphs

Given heterogeneous information network G and subgraph structure set $M = \{M_1, M_2, ..., M_T\}$, we use the network subgraph detection algorithm to match all instances $S_M$ of M on G. If a subgraph structure appears frequently in G, the size of its instances can be reduced based on the strategy of random sampling. Then, the network model instance $S_M$ should be determined. For a network subgraph structure $M_t \in M$, the construction is based on $M_t$ of hypergraph $H_{M_t} = (V_{M_t}, S_{M_t})$, where $S_{M_t}$ are all instances of $M_t$ on G (assuming that the matching instances are not randomly sampled), and $V_{M_t}$ are the set of $S_{M_t}$ covered nodes. Because heterogeneous networks contain rich semantic information, and a single subgraph structure contains only one kind of semantics, it is necessary to consider a variety of high-order subgraph structures to obtain a higher-quality network representation. Therefore, the heterogeneous information network G is transformed into multiple hypergraphs $\{H_{M_1}, H_{M_2}, ..., H_{M_T}\}$, each hypergraph $H_{M_t}$ can be expressed as a binary incidence matrix $H_{M_t} \in R^{N*|S_{M_t}|}$, N is the number of all nodes in the heterogeneous network G, and $|S_{M_t}|$ is the number of hyperedges. Each element in $H_{M_t}$ is defined as:

$$h(v, e) = \begin{cases} 1, & if\ v \in e \\ 0, & if\ v \notin e \end{cases} \quad (1)$$

For a node V, when any super edge $e \in S_{M_t}$ is associated with node V, $h(v, e) = 1$, otherwise it is 0. Each super edge $e \in S_{M_t}$ is assigned a weight $W_e$, and all the weights are stored through the diagonal matrix $W_{M_t} \in R^{|S_{M_t}|*|S_{M_t}|}$. The interaction between different super edges is not considered in this paper. Therefore, $W_{M_t}$ is initialized as the identity matrix here, and all the super edges have the same weight. At the same time, the degree D(V) of node V and the degree $\delta(e)$ of super edge e are stored by diagonal matrices $D_V \in R^{N*N}$ and $D_e \in R^{|S_{M_t}|*|S_{M_t}|}$ respectively. Then the degree of any node is defined as:

$$d(v) = \sum_{e \in S_M} w_e * h(v, e) \quad (2)$$

The number of nodes associated with any superedge e, that is the degree of the superedge, is defined as:

$$\delta(e) = \sum_{v \in V_M} h(v, e) \quad (3)$$

For a given set of network subgraphs M, construct T different adjacency matrices $A = \{A_{M_1}, A_{M_2}, ..., A_{M_T}\}$ based on network subgraphs corresponding to multiple hypergraphs. For any hypergraph $H_{M_t}$, its corresponding adjacency matrix $A_{M_t}$ is defined as:

$$A_{M_t} = H_{M_t} W_{M_t} H_{M_t}^T \quad (4)$$

### 3.2 Dynamic super edge attention mechanism

Graph convolution process is the process of information exchange between adjacent nodes, and the final information distribution reaches balance after several rounds of iterations. In this process, the transfer probability of information between nodes is determined by the adjacency matrix and degree matrix. Hypergraph convolution also constructs a new node representation by multiplying the transition probability of each node with the characteristic matrix, and the transition probability is also determined by the adjacency matrix $A_{M_t}$ and degree matrix $D_v$ and $D_e$ of hypergraph.

Because the transition probability is fixed, the influence of each node on its neighbor nodes will not change in the process of information transmission. Therefore, the super edge internal self attention mechanism is introduced to make the transition probability a trainable parameter, so as to dynamically aggregate different types of nodes. Because the transition probability cannot be parameterized directly, this paper uses the attention mechanism to turn the values in the incidence matrix $H_{M_t}$ into values that can be learned in the process of optimizing the model, so as to overcome the impact of different types of nodes and semantic roles on the representation of target nodes due to their different feature spaces, and indirectly make the model have a trainable transition probability. The new incidence matrix is represented by $\hat{H}_{M_t}$, $a_{ij}$ corresponding to an element in $\hat{H}_{M_t}$. In order to introduce the super edge attention mechanism, the connection probability between different types of nodes can only be 0 or 1, so it is obviously impossible to extract the node information of different semantic roles. The super edge attention mechanism can dynamically assign weights to different types of nodes. For example, when the paper node is the classification target node, different nodes have different influence in the super edge.

For the i-th node, we hope to get the corresponding attention score $a_{ij}$ in the j super edge based on the probability distribution model. Assuming that a certain probability distribution is satisfied for $a_{ij}$, the similarity of node features can be measured uniformly only when the node set comes from the same domain. Therefore, in order to make this calculation feasible, it is necessary to accumulate the characteristics of each node in the super edge to calculate the characteristics of the super edge $e^j$. Here, four accumulation methods are considered: concat, mean, 1-norm, 2-norm, where concat is the feature $\{x_{v^1}, x_{v^2}, ..., x_{v^m}\}$ of each node in $e^j$, m is the number of nodes in the super edge, and

The three methods can be calculated in the following ways:

$$x_{e_n^j} = \begin{cases} \frac{1}{m}(x_{v_n^1} + x_{v_n^2} + ... + x_{v_n^m}) & mean \\ \sqrt[p]{|x_{v_n^1}|^p + |x_{v_n^2}|^p + ... + |x_{v_n^m}|^p} & p-norm \end{cases} \quad (5)$$

where p is equal to 1 or 2, $x_{v_n^i}$ and $x_{e_n^j}$ represent the nth column element of the node and super edge eigenvector respectively. Therefore, referring to the way of calculating attention in the homogeneous network, the super edge self attention mechanism is introduced. For the ith node and its ith super edge, the calculation formula of $a_{ij}$ is

$$a_{i,j} = \frac{\exp(Leaky\,Re\,LU(a[x_{v^i}P_{M_t} \| x_{e^j}P_{M_t}]))}{\sum_{k \in N^j(i)}\exp(Leaky\,Re\,LU(a[x_{v^k}P_{M_t} \| x_{e^j}P_{M_t}]))} \quad (6)$$

Among them, LeakyReLU is a nonlinear activation function. If we need calculate the similarity between heterogeneous nodes, it is necessary to ensure that different types of nodes are in the same vector space, so the weight matrix $P_{M_t}$ is responsible for mapping the eigenvectors $x_{v^i}$ and $x_{e^j}$ to the unified hidden space. ∥ represents the splicing operation. a is a trainable attention vector, which is shared among different hypergraphs. $N^j(i)$ is the neighbor set (including $v^i$) of the node $v^i$ in the super edge. Finally, by learning the self attention in the super edge, the interaction between different types of nodes in the super edge $e^j$ can be described more accurately. In addition, we deeply optimize the hypergraph convolution process by designing self connected parameter $\lambda$ to distinguish the central node from the domain node, replacing the original convolution weight 1 with self connected parameter $\lambda$ to distinguish the self connected edges from the edges between other nodes in the process of graph convolution, so as to enhance the generalization and robustness of the model.

### 3.3 Hypergraph convolution network with integrated semantics

The representation of a single hypergraph $H_{M_t}$ can be obtained based on the above hypergraph convolution network. For the different hypergraph $\{H_{M_1}, H_{M_2}, ..., H_{M_T}\}$ constructed by the network subgraph set M, the network representation $\{Z_{M_1}, Z_{M_2}, ..., Z_{M_T}\}$ with specific semantics is obtained. Because different hypergraphs have different degrees of influence on the generation of the final heterogeneous network representation, the coefficients $C_{M_t}$ calculated based on the attention mechanism are used to fuse the network representation with different semantics

$$C_{M_t} = Soft\max(\tanh(W.Z_{M_t} + b) \cdot a^T) \quad (7)$$

Among them, a is the attention vector used to synthesize semantics, w and b are the weights and offsets of the full connection layer, which is used to project the node representations of different semantics into the same vector space, and then use softmax for logarithmic normalization to obtain the fusion coefficient $C_{M_t}$. Finally, the network representation of each specific semantic is fused to obtain the representation of heterogeneous network G:

$$Z = \sum_{t=1}^{T} C_{M_t} Z_{M_t} \quad (8)$$

For multi-classification tasks, where $Z \in R^{N*q}$, q is the category of nodes. Finally, the loss function of classification is defined as:

$$L = \sum_{i \in V_L} \sum_{j=1}^{q} Y_{ij} \ln Z_{ij} \quad (9)$$

where $V_L$ is the set of all labeled nodes, and Y is the real label of the node. Under the guidance of label information, we can finally get the representation of all nodes and predict their labels based on the parameters in the back-propagation and gradient descent training model.

## 4. Experiments
### 4.1 Datasets

In order to verify the proposed classification method of scientific papers based on Hypergraph neural network, the algorithm is compared with other related algorithms. In addition, we conducted the experiments on three datasets.

**Table 1** Summary of the datasets used in experiments.

|  | DBLP | ACM | AMINER |
|---|---|---|---|
| **Number of nodes** | 18385 | 31636 | 13432 |
| **Number of edges** | 85360 | 162012 | 90387 |
| **Category** | 4 | 3 | 10 |

DBLP dataset: The entities included in this dataset are papers, authors, conferences and teams. Finally, the papers are divided into four categories.
ACM dataset: It contains a large number of scientific papers related to electronic information and computers. And it is divided into three categories. The three categories are database, communication and data mining Through this dataset, we build a heterogeneous map, which includes three types of nodes: paper, author and topic.
Aminer dataset: This dataset includes three node types: author, paper and conference. Then we divide the papers into ten categories.

### 4.2 Baselines

This paper compares the proposed method with the following methods:
DeepWalk[34]: designed for isomorphic graph, based on random walk network;

Metapath2vec: similar to deep walk, skip gram is used to learn the embedding of graphs;

ESIM[35]: Its main idea is to capture semantic information from multiple meta paths;

GCN[36]: It is a semi supervised classification method, mainly used in homogeneous graphs;

GAT[37]: It is a supervised representation learning method, in which attention mechanism is added;

HAN[38]: In order to capture all meta path information, it introduces node level attention mechanism and semantic level attention mechanism;

HGCN[39]: learning node embedding representation using hypergraph neural network model

## 4.3 Experimental results

The main results are shown in the Table 2. SHNN has good performance in three data sets. Among them, SHNN performs best on the ACM data set, and its Macro-F1 index and micro-F1 index are 80.91% and 80.77% respectively. The main reason is that the ACM data set is relatively large and contains more relationships between nodes and edges, which can more comprehensively extract super edges, thus more effectively aggregating the features of related nodes. The last learned node embedding representation is more reasonable.

The experimental results show that the accuracy of SHNN model on three data sets is better than that of other models. In the DBLP and ACM datasets, when the labeling rate is 40%, the classification accuracy is the highest. Moreover, compared with other models, SHNN model can obtain better accuracy from graphs with less labeling proportion, which fully demonstrates the learning ability of this model, so this model can be better applied to graph node classification tasks with less labels.

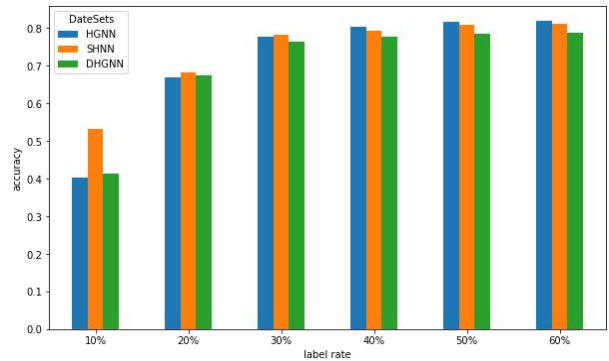

**Figure 2（a）:** Accuracy performance on DBLP

**Table 2:** Experimental results of scientific paper classification.

| Model | DBLP | | | ACM | | | AMINER | | |
|---|---|---|---|---|---|---|---|---|---|
| | Macro-F1 | Micro-F1 | Accuracy | Macro-F1 | Micro-F1 | Accuracy | Macro-F1 | Micro-F1 | Accuracy |
| DeepWalk | 64.17 | 63.88 | 64.12 | 64.81 | 66.26 | 67.12 | 59.12 | 59.52 | 60.15 |
| ESim | 63.01 | 62.89 | 65.11 | 72.53 | 73.44 | 75.11 | 57.26 | 57.96 | 58.21 |
| Metapath2Vec | 63.81 | 63.49 | 64.34 | 71.89 | 72.88 | 75.34 | 68.11 | 68.85 | 69.34 |
| GCN | 63.92 | 63.84 | 63.32 | 72.34 | 73.72 | 76.39 | 68.29 | 69.02 | 71.32 |
| GAT | 76.29 | 76.35 | 75.33 | 80.38 | 80.09 | 80.62 | 69.31 | 69.21 | 71.43 |
| HAN | 77.33 | 77.11 | 77.90 | 79.73 | 79.55 | 79.90 | 70.12 | 70.52 | 72.92 |
| HGNN | 73.48 | 73.26 | 74.89 | 80.81 | 80.69 | 80.59 | 72.26 | 72.01 | 75.59 |
| DHGNN | 78.17 | **78.95** | 78.32 | 80.53 | 80.29 | 89.98 | 74.74 | 74.21 | 77.36 |
| **SPHNN** | **78.23** | 78.22 | **80.56** | **80.91** | **80.77** | **81.61** | **75.25** | **75.58** | **78.56** |

## 4.4 Performance comparison on different labeling rates

In the hypergraph neural network structure, the average classification accuracy of the model is also different due to the different labeling rates of the training data. We tested the change of the average classification accuracy under different datasets when the labeling rate is from 10% to 60%. Specific results are shown in Figure 2.

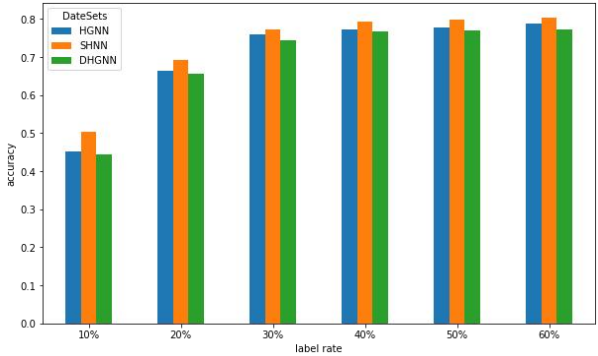

**Figure 2（b）:** Accuracy performance on ACM

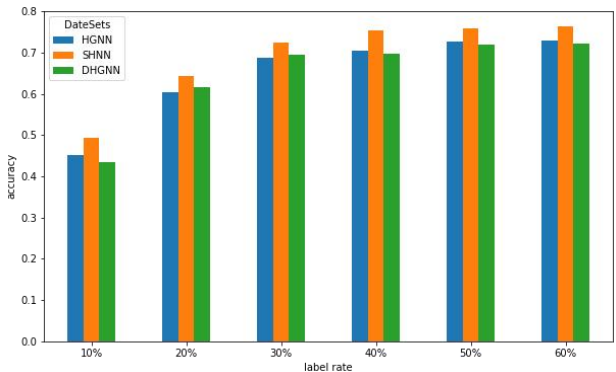

**Figure 2（c）:** Accuracy performance on AMINER

Self connection parameters $\lambda$: We design the self connection parameters $\lambda$ To distinguish the central node from the associated node, use the self connection parameter $\lambda$ Instead of the original convolution weight 1, the self connected edges are separated from the edges of other nodes in the hypergraph convolution process, and the hypergraph convolution process is deeply optimized to enhance the generalization ability and robustness of the model. The following are self connection parameters: $\lambda$ Influence on the experimental training results.

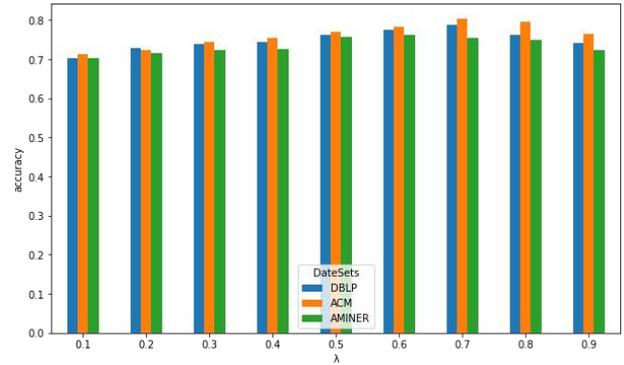

**Figure 3:** Self connection parameters $\lambda$ Influence on model accuracy

It can be seen from the above table that the model has a good effect on the self connection parameters $\lambda$ There is a certain fluctuation in the value change of $\lambda$, when $\lambda$ =0.7, the model achieves the best effect of node classification on DBLP and ACM datasets. And on the AMINER dataset, the optimal corresponding $\lambda$ =0.6。

### 4.5 Ablation study

The importance of ablation research for deep learning research cannot be overemphasized. Understanding causality in a system is the most direct way to generate reliable knowledge (the goal of any research). Ablation is a very simple method to study causality. The SHNN model proposed in this paper consists of two parts, one is the hypergraph construction module, and the other is the attention layer based on graph neural network.we remove each module separately to verify the effectiveness of each part of our proposed model, and then evaluate whether the method proposed by the model has brought about improvement. The final results are shown in Table 3.

**Table 3** Ablation study.

| Model | DBLP | | | ACM | | | AMINER | | |
|---|---|---|---|---|---|---|---|---|---|
| | Macro-F1 | Micro-F1 | Accuracy | Macro-F1 | Micro-F1 | Accuracy | Macro-F1 | Micro-F1 | Accuracy |
| SPHNN-hypergraph | 72.93 | 74.63 | 73.52 | 78.37 | 76.21 | 77.13 | 68.92 | 64.12 | 69.37 |
| SPHNN-attention | 72.73 | 73.62 | 73.14 | 77.86 | 78.77 | 78.26 | 70.61 | 67.68 | 70.62 |
| SPHNN | 78.12 | 79.11 | 79.36 | 80.93 | 81.74 | 81.98 | 75.57 | 73.61 | 75.67 |

According to our experimental results, our method is found to be the best, which proves that these components of the model are effective. By introducing super edge extraction, the correlation between vertex features is extracted to the greatest extent, and the node representation is related to adjacent nodes, that is, to nodes sharing edges. This can be directly applied to inductive learning without acquiring the entire graph. Then, through the attention mechanism of graph neural network, the characteristics of nodes are dynamically aggregated to make the final node representation more accurate.

## 5. Conclusion

This paper proposes a scientific paper classification model based on self-attention and hypergraph. The hypergraph structure is used to aggregate the non-labeled data in the citation relationship of the paper. The hypergraph neural network model based on self-attention realizes the adaptive weight assignment of different depths and different neighbors, which improves the learning ability of the graph neural network model. Experimental results on large-scale real data sets show that the performance of the proposed method is not only better than other traditional graph neural network models, but also better than the traditional supervised text classification model, which proves the effectiveness of the proposed method.

## Acknowledgements

This work was supported by the National Natural Science Foundation of China (No.62192784, No.62172056).